\documentclass[11pt]{article} 
\usepackage{amsfonts,amscd,amsmath}

\def\NP#1#2{ Nucl. Phys. B #1 (#2)} 
\def\PL#1#2{ Phys. Lett. B #1 (#2)}

\def\PRL#1#2{ Phys. Rev. Lett. #1 (#2)} 
\def\PR#1#2{ Phys. Rev. D #1 (#2)} 
\def\IJMP#1#2{ Int. J. Mod. Phys.  A#1 (#2)}

\def\HP#1#2{ JHEP #1 (#2)} 
\def\ap{ \alpha^{\prime}} 

\def\pd{\partial}

\newcommand{\ep}{\text e}
\newcommand{\oh}{\frac{1}{2}}
\newcommand{\qr}{\frac{1}{4}}
\newcommand{\I}{\text I}
\newcommand{\J}{\text J}
\title{Comments on Tachyon Potentials in Closed and Open-Closed String Theories}
\author{Oleg Andreev\thanks{E-mail address: andreev@physik.hu-berlin.de}
\thanks{Also at Landau Institute for Theoretical Physics, Moscow, Russia}
\\ \\
Humboldt--Universit\"at zu Berlin, Institut f\"ur Physik\\
Newtonstra\ss e 15, D-12489 Berlin, Germany}

\date{}
\begin{document} 
 
\maketitle 
\begin{abstract} 
We consider the tachyon potentials in closed and open-closed string theories. In doing so, we apply technique 
which proved to be useful in studying the open string tachyon potentials to the problem of interest. 
\\
PACS : 11.25.-w, 11.25.Pm  \\
Keywords: string theory, tachyons
\end{abstract}

\vspace{-11cm}
\begin{flushright}
hep-th/0308123     \\
HU Berlin-EP-03/46
\end{flushright}
\vspace{9.75cm}

\section{ Introduction} 
\renewcommand{\theequation}{1.\arabic{equation}}
\setcounter{equation}{0}

It is a big problem to better understand the vacuum structure of string/M-theory. In the case of bosonic 
string theory containing the tachyon near its perturbative vacuum, it was realized long time ago that 
the perturbative expansion goes near a wrong vacuum and therefore some sort of spontaneous 
symmetry breaking (the process of condensation of some scalar fields) is needed to generate a true 
vacuum. Numerous attempts have been made to compute the tachyon potentials and find stable minima 
\cite{bar}. Recent years have seen a real progress in our understanding of the problem 
when the open string tachyons were related with annihilation or decay of unstable D-branes via the 
process of their condensation \cite{as, rev}.

In studying the phenomenon of tachyon condensation, string field theory methods are the most 
appropriate ones. However, in practice it turns out to be very hard to deal with them. This is a 
strong motivation for finding simpler methods (toy models)  that can allow one to gain some initial 
intuition on the physics of the phenomenon. For example, the background independent open string 
theory \cite{W,GS,KMM}, p-adic string theory \cite{bf}, and some toy models based on the exactly 
solvable Schr\" odinger problem \cite{z} proved to be useful in studying open string tachyon 
condensation. 

The background independent open string theory is based on the Batalin-Vilkovisky master equation 
whose solution provides the effective action of the theory. In the bosonic case the action is given by

\begin{equation}\label{act}
S_{\oh}=\Bigl(1+\beta^\I\frac{\pd}{\pd g^\I}\Bigr)Z_{\oh}
\quad,
\end{equation}
where $Z_\oh$ is the renormalized partition function of the underlying 2d theory defined on the unit disk 
in the complex plane in such a way that it is a free theory in the bulk but an interacting theory on the 
boundary. $g^\I$ stand for coupling constants of boundary interactions and $\beta^\I$ for their 
RG $\beta$-functions. In the supersymmetric case \footnote{We mean worldsheet SUSY, hear and below.} 
it is simply \cite{a0,kmm,mnp}

\begin{equation}\label{s-act}
S_\oh=Z_\oh
\quad,
\end{equation}

It is worth noting that this theory is closely related with the so-called sigma model approach to string 
theory (see \cite{ts1,ts2} for a review and a list of references). One of the keystones of this approach is the 
condition that the vanishing of the Weyl anomaly coefficients should be equivalent to the conditions of 
stationarity of the effective action. This can be written schematically as

\begin{equation}\label{dif-c}
\frac{\pd S}{\pd g^\I}={\cal G}_{\I\J}\,\bar\beta^\J
\quad,
\end{equation}
where ${\cal G}_{\I\J}$ is a metric on space of the sigma model couplings (background fields) and 
$\bar\beta^\J$ are the Weyl anomaly coefficients which may differ from the $\beta$'s by 
derivative ($\ap$) terms. In fact, all 
constituents of \eqref{dif-c} are defined within the string perturbation theory. In particular, the action is 
represented as $S=\sum_i S_i$, where $i$ is referred to a Riemann surface with $h$ handles and $b$ 
boundaries such that $i=h+\oh b$. So, the expressions \eqref{act}-\eqref{s-act} may be thought of as the 
leading asymptotics of solutions to Eq.\eqref{dif-c} in the case of open strings.

At present, the leading asymptotics from the string massless modes and the open string tachyons are relatively 
good understood. On the other hand, there is no clear understanding of what happens with the closed 
string tachyons. As noted above, there are the methods which turned out to be useful in studying the 
open string tachyons. So, our idea here is to try these to describe effective actions with the closed string 
tachyons.

The paper is organized as follows. We begin with the p-adic string theory including both open and closed 
string tachyons in section 2. The potential of interest is obtained by taking the limit $p\rightarrow 1$. In 
section 3, we compute the potential by solving Eq.\eqref{dif-c} and compare it with the result of section 2. The 
potentials are the same up to one term which is crucial for missing of a close string tachyon tadpole in the 
potential obtained from the p-adic theory. Such a remarkable coincidence provides a nontrivial check of 
the universality of our results. Section 4 will present a generalization to type 0 theories with the closed 
string tachyon coming from the $(NS-,NS-)$ sector and some open problems.

\section{A first look at the open-closed tachyon potential} 
\renewcommand{\theequation}{2.\arabic{equation}}
\setcounter{equation}{0}
There have been some recent attempts to understand open as well as closed tachyon condensations within 
the p-adic string theory \cite{z1, ms}. As known, these models are useful in studying open string tachyon 
condensation, where it is easy to write down the tachyon potential by taking the 
limit $p\rightarrow 1$ in the corresponding p-adic theory \cite{GS}. In this section, we extend 
the analysis of \cite{GS} to include the effects of the closed string tachyon. Although the p-adic string 
theory including the closed string tachyon suffers from some drawbacks, we belive, nevertheless, that 
we can gain some insight on the effects of the closed string tachyon. We will return to this issue in 
section 3.

\subsection{Tachyon action from p-adic strings}
The action that reproduces p-adic tachyon amplitudes is given by \cite{bf}

\begin{equation}\label{p-ac}
\begin{split}
S=\int d^D x\,\biggl[\,\, 
&\frac{1}{g^2_c}\biggl(-\oh\frac{p^4}{p^2-1}\,\phi \,p^{-\ap\Box /2}\,\phi
+\frac{p^4}{p^4-1}\,\phi^{p^2+1}\biggr)\\
+&\frac{1}{g^2_o}\biggl(-\oh\frac{p^2}{p-1}\,\Phi \,p^{-\ap\Box }\,\Phi 
+\frac{p^2}{p^2-1}\,\phi^{p(p-1)/2}\,\bigl(\Phi^{p+1}-1\bigr)\biggl)
\,\,\biggr]
\quad,
\end{split}
\end{equation}
 where $\phi$ and $\Phi$ are the closed and open string tachyons, $g_c$ and $g_o$ are their 
coupling constants. 

The equations of motion are 

\begin{equation}\label{p-eq}
-p^{-\ap\Box}\,\Phi+\phi^{p(p-1)/2}\,\Phi^p=0
\quad,
\quad
-p^{-\ap\Box/2}\,\phi+\phi^{p^2}+\frac{g}{2}\frac{p-1}{p}\,
\phi^{p(p-1)/2-1}\bigl(\Phi^{p+1}-1\bigr)=0
\quad,
\end{equation}
with $g=g^2_c/g^2_o$.

If we take the limit $p\rightarrow 1$, then we find to leading non-trivial order in $p-1$

\begin{equation}\label{t-eq}
\ap\Box\,\Phi+\oh\bigl(\ln\Phi^2 +\oh\ln\phi^2\bigr)\Phi=0
\quad,
\quad
\ap\Box\,\phi+2\,\phi\,\ln\phi^2+g\,\frac{\Phi^2-1}{\phi}=0
\quad.
\end{equation}
These equations allow us to write down the desired action \footnote{Our sign convention 
is $\eta_{\mu\nu}=\text{diag}(-1,1,\dots ,1)$.}

\begin{equation}\label{t-ac}
S=\int d^D x\biggl[\,\, 
\frac{1}{g^2_c}
\biggl(\frac{1}{8}\ap\pd_\mu\phi\pd^\mu\phi-\frac{1}{4}\,\phi^2\ln\frac{\phi^2}{\ep}\biggr)
+\frac{1}{g_o^2}
\biggl(\oh\ap\pd_\mu\Phi\pd^\mu\Phi-\frac{1}{4}\,\Phi^2\ln\frac{\Phi^2}{\ep}
-\frac{1}{8}\bigl(\Phi^2-1\bigr)\ln\phi^2 \biggr)
\,\,\biggr]
\quad.
\end{equation}

It will be useful for the following to introduce new fields as $\phi=\ep^{-t/2}$ and $\Phi=\ep^{-T/2}$. In 
terms of these (modulo an overall factor), the action takes the form

\begin{equation}\label{T-ac}
S=\int d^D x\biggl[\,\, 
\frac{1}{g^2_c}\,\ep^{-t}
\biggl(\frac{1}{8}\ap\pd_\mu t\pd^\mu t+1+t \biggr)
+\frac{1}{g_o^2}\,\ep^{-T}
\biggl(\oh\ap\pd_\mu T\pd^\mu T+1+T+\oh\, t-\oh\, t\ep^T\biggr)
\,\,\biggr]
\quad.
\end{equation}

\subsection{Analysis of the potential}

Having derived the action, let us now explore the potential. It is given by 

\begin{equation}\label{T-p}
V=-\frac{1}{4g^2_c}\,\phi^2\ln\frac{\phi^2}{\ep}-
\frac{1}{4g^2_o}\,\biggl(\Phi^2\ln\frac{\Phi^2}{\ep}+\oh\bigl(\Phi^2-1\bigr)\ln\phi^2\biggr)
\quad.
\end{equation}

It is a trivial exercise to show that the potential is stationary at three points of the 
$\phi\text{\small -}\Phi$ plane: \footnote{We consider non-negative values of $\phi$ and $\Phi$, here 
and below.}

(i) $\phi=1,\,\,\Phi=1$. This point is a local maximum which represents the standard 
perturbative vacuum of the model.
 
(ii) $\phi=\phi_o,\,\,\Phi=0$, where $\phi_o$ is a solution to $2\phi^2\ln\phi^2=g$. This is a 
saddle point. Excitations of $\Phi$ are completely suppressed while excitations of $\phi$ develop 
instability. So, it can be interpreted as a closed string background without D-branes. The rolling from the 
perturbative vacuum to this point is open string tachyon condensation. The model allows us to estimate 
some consequences of the process for the closed string tachyon (closed string background). For instance, its 
vacuum expectation value and mass get shifted: $\langle\phi\rangle =1\rightarrow 
\langle\phi\rangle =\phi_o=\sum_{n=0}^\infty a_n\,g^n=1+\frac{1}{4}g+O(g^2)$, 
$m_c^2=-\frac{4}{\ap} \rightarrow m_c^2=-\frac{4}{\ap}\bigl(1+\frac{g}{2\phi_o^2}\bigr)$.

(iii) $\phi=\phi_c,\,\,\Phi=1/\sqrt\phi_c$, where $\phi_c$ is a solution to 
$2\phi^3\ln\phi^2=g(\phi -1)$. This is another saddle point. There are excitations of $\Phi$ as well 
as $\phi$. Instability is now due to $\Phi$. Therefore it seems natural to interpret this point as a closed 
string background with D-branes. The rolling from the perturbative vacuum to this point is 
closed string tachyon condensation. In contrast with $\phi_o$, $\phi_c$ doesn't admit a series 
expansion in $g$. The effect on the open string tachyon is huge: its vev gets shifted from 1 
to $1/\sqrt{\phi_c}$ which is very large as $\phi_c\rightarrow 0$ for $g\rightarrow 0$. 

Let us conclude this section by making a couple of remarks.

(a) As we have just seen, the potential doesn't possess a local minimum which could be 
interpreted as a true vacuum.\footnote{Note that for $p\geq 2$ a local minimum
always exists \cite{ms}.} We do not think, however, that this is a real disaster. The point is that the 
potential under consideration is not bounded from below. So, given a local minimum (with a finite value of $V$), 
this is a false vacuum because of tunneling through the potential barrier.\footnote{For a discussion 
of this issue in the case of the open string tachyon alone, see \cite{a1}.} A more crucial point is that the 
model at hand is just the theory of two scalar fields rather than string field theory or, at least, 
the theory of the scalars interacting with stringy massless modes. The effects of these modes are not 
always negligible. For example, gravitation is sometimes important for the decay process of false 
vacua \cite{CL}. We will return to this point in section 3. 

(b) While the first two terms of the potential are symmetric under a permutation of 
$\bigl(\phi,\,g_c\bigr)$ with $\bigl(\Phi,\,g_o\bigr)$, the third term mixing the closed and open 
string tachyons completely destroys this symmetry. As a result, there is a drastic difference between the 
processes of closed and open string tachyon condensation, i.e., between (ii) and (iii). Moreover, we will 
see in section 3 that the form of this term is also crucial for the theory to have a local minimum.

\section{Partition function representation for closed bosonic string theory effective action with tachyon} 
\renewcommand{\theequation}{3.\arabic{equation}}
\setcounter{equation}{0}

We have found the tachyon potential, Eq.\eqref{T-p}, by taking the limit $p\rightarrow 1$ in 
the p-adic theory. We believe, however, that the result is rather universal. In order to illustrate 
this, we will study the partition function representation of the effective action as it was done in 
\cite{W,GS,KMM} for the open string tachyon.

\subsection{A warm-up example: constant closed string backgrounds}

Consider the standard sigma model action 

\begin{equation}\label{sigma}
I=\frac{1}{4\pi\ap}\int_{S^2} d^2z\sqrt{g}\biggl[ \pd_aX^\mu\pd^a X^\nu {\mathbf G}_{\mu\nu}(X)+
\ap\,{\mathbf t}(X)+\ap R^{(2)}\boldsymbol{\varphi}(X)\biggr]
\end{equation}
defined on a two-dimensional sphere of radius $r$. In the case of constant background fields 
it can be shown that the corresponding bare partition function is given by 
\cite{ft} \footnote{For the sake of simplicity, 
we discard all numerical constants which accompany the $\ln\frac{\mu}{\Lambda}$'s, here 
and below.}

\begin{equation}\label{pfun}
\mathbf{Z}_0=z_0\int d^Dx\,\sqrt{\mathbf{G}}\,\ep^{-2\boldsymbol{\varphi}-
r^2\mathbf{t}-\lambda\ln\frac{\mu}{\Lambda}}
\quad,
\end{equation}
where $z_0$ is a normalization constant, $\lambda=(D-26)/3$, $\mu$ is a scale of the world-sheet 
theory such that $r=1/\mu$, $\Lambda$ is a momentum space UV cut-off. By using the following 
subtraction to renormalize the couplings 

\begin{equation}\label{ren}
\boldsymbol{\varphi}=\varphi-\oh\lambda\ln\frac{\mu}{\Lambda}
\quad,\quad
\mathbf{G}_{\mu\nu}=G_{\mu\nu}
\quad,\quad
\mathbf{t}=\mu^2\,t
\end{equation}
the renormalized partition function takes the form

\begin{equation}\label{pfun1}
Z_0=z_0\int d^Dx\,\sqrt{G}\,\ep^{-2\varphi-t}
\quad.
\end{equation}

Note that the expressions \eqref{ren} also determine the RG $\beta$-functions

\begin{equation}\label{beta}
\beta^\varphi=\oh\lambda
\quad,\quad
\beta^G_{\mu\nu}=0
\quad, \quad
\beta^t=-2t
\quad.
\end{equation}

In our special case, the partition function and action depend on the space-time metric 
$G_{\mu\nu}$ only through a volume factor $V_D=\int d^Dx\,\sqrt{G}$. Thus, all that remains 
is to determine $S_0/V_D$ as a function of 
$t$ and $\varphi$. To this end, we make a field redefinition: $(t,\varphi )\rightarrow (\tau,\varphi)$, 
where $\tau=t+2\varphi$. This leads to $Z_0/V_D=z_0\,\ep^{-\tau}$ and 
$\beta^\tau=-2\tau+\lambda+4\varphi$. The metric on field space now has the only 
component ${\cal G}_{\tau\tau}=z_0\,\ep^{-\tau}$.\footnote{As is usual, the metric is defined by 
${\cal G}_{\I\J}=\pd_\I\pd_\J Z$, where $\pd_\I=\pd /\pd g^\I$.} According to \eqref{dif-c},  $S_0/V_D$ 
obeys the equation

\begin{equation}\label{cl}
\frac{\pd}{\pd\tau} S_0/V_D=z_0\,\ep^{-\tau}\bigl(-2\tau+\lambda+4\varphi\bigr)
\end{equation}
that is easily integrated to 

\begin{equation}\label{ac}
S_0/V_D=z_0\,\ep^{-\tau}\bigl(2+2\tau-\lambda-4\varphi\bigr)+f_0(\lambda,\varphi)
\quad.
\end{equation}
Here $f_0$ is an integration ``constant'' which is a function of $\lambda$ and $\varphi$.

We can now write down the effective action

\begin{equation}\label{ac2}
S_0=z_0\int d^Dx\,\sqrt{G}\biggl[\ep^{-2\varphi-t}\bigl(2+2t-\lambda\bigr)
+f_0(\lambda,\varphi)\biggr]
\quad.
\end{equation}
$f_0$ and $z_0$ are determined by the condition that for $t=0$ the action reduces to the well-known 
result of \cite{callan,Jb}

\begin{equation*}
S_0=\frac{1}{\kappa_0^2\ap}\,\int d^Dx\,\sqrt{G}\ep^{-2\varphi}\,\lambda
\quad.
\end{equation*}
We find
\begin{equation}\label{f}
f_0(\lambda,\varphi)=2(\lambda-1)\,\ep^{-2\varphi}
\quad,\quad
z_0=\frac{1}{\kappa^2_0\ap}
\quad.
\end{equation}

Finally, the effective action takes the form

\begin{equation}\label{ac3}
S_0=\frac{2}{\kappa_0^2\ap}
\int d^Dx\,\sqrt{G}\,\ep^{-2\varphi}\Bigl[\ep^{-t}\bigl(1+t-\oh\lambda\bigr)+\lambda-1\Bigr]
\quad.
\end{equation}
At this point a few remarks are in order:

(a) We now see that after shifting $t$ by $t\rightarrow t+\oh\lambda$ the closed string tachyon 
potential following from \eqref{ac3} takes the same form as that derived from the p-adic 
string.\footnote{To compare them, we set $T=0,\,\varphi=\varphi_0,\,G_{\mu\nu}=\eta_{\mu\nu},\,
g_c^2=\oh\kappa^2_0\ap\ep^{2\varphi_0}$.} This provides a nontrivial check of the 
universality of our results. Note that the $\lambda$-term in \eqref{ac3} is the contribution of the 
dilaton field which is absent in the p-adic theory.

(b) We may also rewrite Eq.\eqref{ac3} as 

\begin{equation}\label{ac4}
S_0=\Bigl(2+\beta^\I\frac{\pd}{\pd g^\I}\Bigr)Z_0+\int d^Dx\,\sqrt{G}\,f_0
\quad.
\end{equation}
Thus, a principal difference between this expression and that of \eqref{act} is the integration ``constant'' 
$f_0$. We now understand what is happening. 
The mixing of the tachyon with the dilaton takes place and, as a consequence, there is the only 
differential equation which defines the effective action. A closely related fact is that the standard 
perturbative vacuum $(t=0,\,\lambda=0,\,\varphi=const)$ is degenerate: it is a line 
in the $t\text{\small -}\varphi$ plane. 

A factor of $2$ as compared to \eqref{act} keeps track of the closed string tachyon dimension which is 
twice as big as the open string tachyon dimension.

(c) Alternatively, we can get the action \eqref{ac2} by integrating

\begin{equation}\label{t-e}
\frac{\pd S_0}{\pd t}={\cal G}_{tt}\,\beta^t +{\cal G}_{t\varphi}\,\beta^\varphi+
{\cal G}_{tG}^{\mu\nu}\,\beta^G_{\mu\nu}
\quad,
\end{equation}
with ${\cal G}_{tt}=z_0V_D\,\ep^{-2\varphi-t}$, ${\cal G}_{t\varphi}=2\,{\cal G}_{tt}$, and 
${\cal G}_{tG}^{\mu\nu}=-\oh\,{\cal G}_{tt}G^{\mu\nu}$.

(d) A final remark: it was proposed by Tseytlin in \cite{ts-m} that the effective action for the closed 
string massless modes is simply $S\sim\beta^\I\pd_\I Z$. One can think of $2Z$ and $f_0$ as a 
further refinement of this definition for massive modes. This is quite opposite to what happened in the 
case of open string, where the action of its massless mode is $Z$ and $\beta^\I\pd_\I Z$ is a further 
refinement for massive modes.

\subsection{Leading $\ap$-corrections}
We now turn to the problem of the evaluation of $\ap$-corrections to the effective 
action \eqref{ac3}. To do so, we need to solve Eq.\eqref{dif-c} to next order in $\ap$. Since the sigma 
model is renormalizable within the $\ap$-expansion, one may expect that the metric on field space and 
the $\bar\beta$'s are given by 

\begin{equation}\label{ap}
{\cal G}_{\I\J}={\cal G}_{\I\J}^{(0)}+\ap\,{\cal G}_{\I\J}^{(1)}+O(\ap\,^2)
\quad,\quad
\bar\beta^\I=\bar\beta^{\I\,(0)}+\ap\,\bar\beta^{\I\,(1)}+O(\ap\,^2)
\end{equation}
that results in \footnote{For non-constant backgrounds, $\frac{\pd}{\pd g^\I}$ is 
promoted to $\frac{\delta}{\delta g^\I}$.}

\begin{equation}\label{ap1}
\frac{\delta S}{\delta g^\I}={\cal G}_{\I\J}^{(0)}\cdot\bar\beta^{\J\,(0)}+
\ap\,\Bigl({\cal G}_{\I\J}^{(1)}\cdot\bar\beta^{\J\,(0)}+{\cal G}_{\I\J}^{(0)}\cdot\bar\beta^{\J\,(1)}
\Bigr)+O(\ap\,^2)
\quad.
\end{equation}
Thus, a naive comparison between the vanishing of the Weyl anomaly coefficients and the conditions 
of stationary of the action might be confusing at this level. In this paper we will not work this out in full detail, 
but just carry out some calculations in two different schemes. 

The bare partition function is given by 

\begin{equation}\label{pf-ap}
\mathbf{Z}_0=z_0\int d^Dx\,\sqrt{\mathbf{G}}\,
\ep^{-2\boldsymbol{\varphi}-r^2\mathbf{t}-\lambda\ln\frac{\mu}{\Lambda}}
\biggl[1+
\ap\ln\frac{\mu}{\Lambda}
\Bigl(D^2\boldsymbol{\varphi}+\oh D^2\mathbf{t}+\oh\mathbf{R}\Bigl)+O(\ap\,^2)
\biggr]
\quad.
\end{equation}
We chose a special subtraction scheme

\begin{equation}\label{scheme}
\boldsymbol{\varphi}=\varphi-\oh\bigl(\lambda-\ap D^2\varphi\bigr)\ln\frac{\mu}{\Lambda}
\quad,\quad
\mathbf{G}_{\mu\nu}=G_{\mu\nu}-\ap R_{\mu\nu}\ln\frac{\mu}{\Lambda}
\quad,\quad
\mathbf{t}=\mu^2\Bigl(t+\oh\ap D^2t\ln\frac{\mu}{\Lambda}\Bigr)
\end{equation}
such that the renormalized partition function takes the form

\begin{equation}\label{pf-ap1}
Z_0=z_0\int d^Dx\,\sqrt{G}\,
\ep^{-2\varphi-t}
\quad.
\end{equation}

From \eqref{scheme} we can also reproduce the standard $\beta$-functions of the fields

\begin{equation}\label{beta-1}
\beta^\varphi=\oh\lambda-\oh\ap D^2\varphi
\quad,\quad
\beta^G_{\mu\nu}=\ap R_{\mu\nu}
\quad, \quad
\beta^t=-2t-\oh\ap D^2 t
\quad
\end{equation}
which differ from the corresponding Weyl anomaly coefficients by ``diffeomorphism'' terms
\begin{equation}\label{Weyl}
\bar\beta^\varphi=\beta^\varphi+\ap\,\pd^\mu\varphi\pd_\mu\varphi
\quad,\quad
\bar\beta^G_{\mu\nu}=\beta^G_{\mu\nu}+2\ap\,D_\mu D_\nu\varphi
\quad, \quad
\bar\beta^t=\beta^t +\ap\,\pd^\mu\varphi\pd_\mu t
\quad.
\end{equation}
It is a trivial exercise to show that in our approximation $\bar\beta^\I\cdot
\frac{\delta Z_0}{\delta g^\I}=\beta^\I\cdot\frac{\delta Z_0}{\delta g^\I}$.

We are ready to use \eqref{ac4}.\footnote{We might remark at this point that $f_0$ actually is 
$-2\bigl(\beta^\I\cdot\frac{\delta Z_0}{\delta g^\I} +Z_0\bigr)$ at $t=0$.} The result is  

\begin{equation}\label{ap-a}
\begin{split}
S_0=&\frac{1}{2\kappa_0^2}
\int d^Dx\,\sqrt{G}\,\ep^{-2\varphi}
\biggl[\,\frac{4}{\ap}\,\bigl(\lambda -1\bigr)-2\,R-8\,\pd_\mu\varphi\pd^\mu\varphi
\\
&
+\ep^{-t}\,\Bigl(\frac{4}{\ap}\,\bigl(1+t-\oh\lambda\bigr)+
R+4\,\pd_\mu\varphi\pd^\mu\varphi+4\,\pd_\mu\varphi\pd^\mu t+
\pd_\mu t\pd^\mu t+\dots \Bigr)
\biggr]
\quad,
\end{split}
\end{equation}
where the ellipsis stands for cubic terms in $t$ and $\varphi$. 

Note that as in the open string case $\pd t\pd t$ is non-universal in the sense that its coefficient 
is renormalization scheme dependent. The universal thing to compute is, for instance, 
the mass of the tachyon in the perturbative vacuum that comes from all derivative terms.\footnote{For a 
discussion of this issue in the case of the open string tachyon see, e.g., \cite{KMM}.} We use the special scheme 
which turns out to be more convenient for practical calculations dealing with $\ap$-corrections in 
subsection 3.4. It was also used for similar purposes in the background independent open string 
theory \cite{W,GS,KMM}. A general form would be $c_1\pd t\pd t$.  Another way to fix $c_1$ is the 
following. Varying $t$, 

\begin{equation}\label{t}
\frac{\delta S_0}{\delta t}=\frac{1}{\kappa_0^2\ap}\sqrt{G}\,\ep^{-2\varphi}
\Bigl[-2t-\ap c_1D^2t+\lambda-\ap D^2\varphi-
\oh\ap G^{\mu\nu}\bigl(R_{\mu\nu}+2 D_\mu D_\nu\varphi\bigl)\Bigr]
\quad.
\end{equation}
In the bracket we keep only the linear terms in $t$ and $\varphi$. Next assuming 
${\cal G}^{(1)}_{tt}=0$, 
we get that for $c_1=1/2$ the equation of motion is consistent with the Weyl anomaly 
coefficients \eqref{Weyl}.

Note that the choice $c_1=1$ seems more natural because on general grounds the kinetic terms should 
be given by ${\cal G}_{\I\J}\cdot\pd g^\I\pd g^\J$. Then noting  

\begin{equation}\label{G}
{\cal G}^{(0)}_{tt}(x,y)=z_0\,\ep^{-2\varphi-t}\delta^{(D)}(x-y)
\quad,\quad
{\cal G}^{(0)}_{t\varphi}(x,y)=2\,{\cal G}^{(0)}_{tt}(x,y)
\end{equation}
immediately leads to $c_1=1$. Contrary to the case of $c_1=1/2$ the leading 
$\ap$-correction to ${\cal G}^{(0)}_{tt}$ is now non-zero. 

For future reference, we note that for $D=26$ we can also write the action \eqref{ap-a} near the 
perturbative vacuum ($t\sim 0$) as follows. First, we make the standard Weyl rescaling that brings 
the action into the form 

\begin{equation}\label{a26}
\begin{split}
S_0=&\frac{1}{2\kappa^2}
\int d^{26}x\,\sqrt{\tilde G}\,
\biggl[\,-\frac{4}{\ap}\ep^{\tilde\varphi/6}-2\,\tilde R
+\frac{1}{3}\,\pd_\mu\tilde\varphi\pd^\mu\tilde\varphi
\\
&
+\ep^{-t}\,\Bigl(\frac{4}{\ap}\,\bigl(1+t\bigr)\ep^{\tilde\varphi/6}+
\tilde R-\frac{1}{6}\,\bigl(\pd_\mu\tilde\varphi\pd^\mu\tilde\varphi+
\pd_\mu\tilde\varphi\pd^\mu t\bigr)+\pd_\mu t\pd^\mu t\Bigr)
\biggr]
\quad,
\end{split}
\end{equation}
where $\varphi=\varphi_0+\tilde\varphi ,\,G_{\mu\nu}=\ep^{\tilde\varphi/6}\tilde G_{\mu\nu}
\,,\,\,\kappa=\kappa_0\ep^{\varphi_0}$. Expanding then the exponents in powers of the fields 
and keeping the leading terms, we get 
\begin{equation}\label{a26-1}
S_0=\frac{1}{2\kappa^2}
\int d^{26}x\,\sqrt{\tilde G}
\biggl[\,\pd_\mu t\pd^\mu t+\oh m^2\,t^2-\tilde R+
\frac{1}{6}\,\pd_\mu\tilde\varphi\pd^\mu\tilde\varphi+\Bigl(\frac{1}{6}D^2\tilde\varphi-
\tilde R\Bigr)t
\biggr]
\quad,
\end{equation}
where $m^2=-4/\ap$. At this point we redefine the tachyon field as 
$\tilde t=t+\qr\ap\bigl(\tilde R-\frac{1}{6}D^2\varphi\bigr)$ to make the kinetic matrix diagonal. 
The action is 

\begin{equation}\label{a26-2}
S_0=\frac{1}{2\kappa^2}
\int d^{26}x\,\sqrt{\tilde G}
\biggl[\,\pd_\mu\tilde t\pd^\mu \tilde t+\oh m^2\,\tilde t^2-\tilde R+
\frac{1}{6}\,\pd_\mu\tilde\varphi\pd^\mu\tilde\varphi
\biggr]
\quad.
\end{equation}
One thing about \eqref{a26-2} may be disturbing. It seems that the tachyon mass is 
given by $m^2=-2/\ap$ instead of $m^2=-4/\ap$. The answer to this puzzle is known \cite{KMM}. The 
$\ap$-expansion assumes zero momenta while the tachyon on shell is far from zero momenta. So, the 
higher derivative corrections become important for going on shell. Loosely speaking, their effect should be a
multiplicative renormalization $m^2\rightarrow 2m^2$ in Eq.\eqref{a26-2}. As a result, the 
effective action with the correct mass-shell conditions for the fields becomes

\begin{equation}\label{a26-3}
S_0=\frac{1}{2\kappa^2}
\int d^{26}x\,\sqrt{\tilde G}
\biggl[\,\pd_\mu\tilde t\pd^\mu \tilde t+m^2\,\tilde t^2-\tilde R+
\frac{1}{6}\,\pd_\mu\tilde\varphi\pd^\mu\tilde\varphi
\biggr]
\quad.
\end{equation}

\subsection{A brief analysis of the action}

Given the action \eqref{ap-a}, it is straightforward to show that the tachyon potential is 
stationary at $t=\oh\lambda$ and $t=\infty$. For $D=26$, the former is the standard perturbative 
vacuum, near which $V(t)=-\frac{2}{\ap }t^2+O(t^3)$. There is no cosmological constant, so the 
spacetime geometry is a $26$-dimensional Minkowski space with a constant dilaton, as we would 
expect. The ``stable vacuum'' to which the 
tachyon might condense is at $t=\infty$.\footnote{We don't consider the possibility for the tachyon instead 
to roll down to $t=-\infty$.} In this vacuum the tachyon disappears from the spectrum by developing an 
infinite mass. An interesting consequence of tachyon condensation is a non-zero cosmological constant. 
Its value can be determined by noting that at $t=\infty$ the action reduces to 

\begin{equation}\label{ac-in}
S_0=\frac{1}{\kappa_0^2}
\int d^{26}x\,\sqrt{G}\,\ep^{-2\varphi}
\Bigl[\,-\frac{2}{\ap}-\,R-4\,\pd_\mu\varphi\pd^\mu\varphi\Bigr]
\quad.
\end{equation}
It is clear that $G_{\mu\nu}=\eta_{\mu\nu},\,\varphi=const$ is not an extremum of 
the action \eqref{ac-in}. Instead we should look for solutions that are more complicated than 
Minkowski space. 

Before continuing our discussion, it's worth noting that such a scenario for generating a non-zero 
cosmological constant is an old idea. For instance, in the context of string cosmology it was discussed 
by Kostelecky and Perry \cite{perry}. In fact, we can obtain the static solutions we are looking for 
from the cosmological solutions of \cite{perry} by continuation 
$y\rightarrow ix^0,\,x^0\rightarrow iy$. Let us consider $SO(1,d)$ invariant solutions of the form

\begin{equation}\label{sol}
ds^2=dy^2+a^2(y)\eta_{ij}dx^idx^j
\quad,\quad
\varphi=\varphi(y)
\quad,
\end{equation}
where $i,j=0,\dots ,d$ and $d\leq D-2$. For convenience we have given these for general $d$ and 
$D$. The scale factor and the dilaton are given by 

\begin{equation}\label{sol1}
a(y)=a_0\bigl(\tanh Ay\bigr)^\Delta
\quad,\quad
\varphi(y)=\oh\bigl(d\Delta-1\bigr)\ln\sinh Ay -\oh\bigl(d\Delta+1\bigr)\ln\cosh Ay\,+const
\quad,
\end{equation}
with $d\Delta^2=1,\,A^2=(1-\lambda)/2\ap$. 

It is of some interest to consider special asymptotics of the solutions. For large positive $y$, the 
solutions behave as 
\begin{equation}\label{sol2}
a=const
\quad,\quad
\varphi=-Ay
\quad,
\end{equation}
while for small positive $y$, 
\begin{equation}\label{sol3}
a\sim y^\Delta
\quad,\quad
\varphi\sim\ln y
\quad.
\end{equation}
The former is the well-known linear dilaton background which corresponds to CFT on 
the world-sheet. This provides some evidence that we have indeed found the vacuum state. As to 
the latter, it is unclear how 
to relate it to world-sheet CFT. Another trouble is that the corresponding curvature becomes large, so 
higher derivative corrections might be relevant in this limit. What really happens is presently unknown. 
Note, however, that a similar asymptotic behavior was discussed by Polyakov in the context of gravity coupled 
to conformal matter with $c>1$  as a possible solution to the so-called problem of $c=1$ barrier \cite{am}.

A final important remark: the study of open string tachyon condensation provided strong evidence that 
all open string modes disappear from the spectrum as the open string tachyon condenses. For instance, 
one indication is that the metric on field space being proportional to $\ep^{-T}$ degenerates as 
$T\rightarrow\infty$. In contrast, closed string modes can survive during the process of closed string 
tachyon condensation. Indeed, the kinetic terms of the graviton and the dilaton don't vanish at $t=\infty$ 
as seen from \eqref{ap-a}. As a result, the metric is not completely degenerate.

\subsection{Free field background}

There are some motivations to consider free field backgrounds. First, the theory based on such a 
background is exactly solvable as the world-sheet path integral remains Gaussian. Second, it allows 
one to effectively resum some $\ap$-corrections and somewhat to go beyond the $\ap$-expansion 
with its restrictive range of validity. Finally, the 
free field backgrounds turned out to be useful tool in studying open string tachyon 
condensation \cite{W,GS,KMM}. Our hope is that these might also be useful for the problem of interest. 

A particularly simple background to be considered is 

\begin{equation}\label{back}
\mathbf{t}(X)=\mathbf{t}_0+\sum_i\Bigl(2\mathbf{w}_iX_i+\mathbf{u}_iX_i^2\Bigr)
\quad,\quad
\boldsymbol{\varphi}(X)=\boldsymbol{\varphi}_0+\sum_i\mathbf{v}_iX_i
\quad,\quad
\mathbf{G}_{ij}=\delta_{ij}
\quad,
\end{equation}
where $i=1,\dots ,d$.

The bare partition function is found by expanding the $X$'s in the spherical harmonics and 
performing the corresponding Gaussian integrals. The result is 

\begin{equation}\label{2pf}
\mathbf{Z}_0=z_0\,\tilde V_{d}\,
\ep^{-\lambda\ln\frac{\mu}{\Lambda}-2\boldsymbol{\varphi}_0-r^2\mathbf{t}_0}
\,
\prod_i \frac{1}{\sqrt{r^2\mathbf{u}_i}}\,
\ep^{\frac{\mathbf{s}_i^2}{r^2\mathbf{u}_i}}
\,\prod^\infty_{n=1}\biggl[1+\ap\,\frac{r^2\mathbf{u}_i}
{n^2+n}\biggr]^{-n-\oh}
\quad,
\end{equation}
where $\mathbf{s}_i=r^2\mathbf{w}_i+\mathbf{v}_i$, $\tilde V_{d}=\int d^{D-d}x$.

The expression \eqref{2pf} is pure formal because it is divergent. To proceed further, we 
regularize the exact propagator of the $X$'s at the coinciding points 
as \footnote{Some useful formulae can be found in \cite{amts}.}

\begin{equation}\label{reg}
\langle X^i X^j\rangle= -\oh\ap\delta^{ij}\sum_{n=0}^\infty 
\frac{2n+1}{n^2+n+\ap\,r^2\mathbf{u}_i}\,\ep^{-n(n+1)\epsilon^2}
\quad,
\end{equation}
where $\epsilon=\mu/\Lambda,\,\mu=1/r$. The preceding calculation of $\mathbf{Z}_0$ can be generalized 
without difficulty to this case. The partition function takes the form

\begin{equation}\label{2pf0}
\mathbf{Z}_0=z_0\,\tilde V_d\,
\ep^{-\lambda\ln\epsilon-2\boldsymbol{\varphi}_0-r^2\mathbf{t}_0}
\prod_i \frac{1}{\sqrt{r^2\mathbf{u}_i}}\,\ep^{\mathbf{F}_i}
\quad,
\end{equation}
with 
\begin{equation*}
 \mathbf{F}_i=\frac{\mathbf{s}_i^2}{r^2\mathbf{u}_i}+\ap\,\Bigl(\ln\epsilon
+\oh\bigl(1-\gamma\bigr)\Bigr)r^2\mathbf{u}_i+
\oh\sum_{m=2}^\infty \frac{(-1)^m}{m}\,\bigl(\ap\,r^2\mathbf{u}_i\bigr)^m
\sum^\infty_{n=1}\frac{2n+1}{(n^2+n)^m}\,+O(\epsilon)
\quad.
\end{equation*}
Here $\gamma$ denotes the Euler's constant.

We renormalize the couplings via the minimal subtraction

\begin{equation}\label{re}
r^2\mathbf{t}_0=t_0+\ap\ln\epsilon\sum_i u_i
\,\,,\quad
r^2\mathbf{w}_i=w_i
\,\,,\quad
r^2\mathbf{u}_i=u_i
\,\,,\quad
\boldsymbol{\varphi}_0=\varphi_0-\oh\lambda\ln\epsilon
\,\,,\quad
\mathbf{v}_i=v_i
\,\,.
\end{equation}
This implies that the corresponding $\beta$-functions are given by

\begin{equation}\label{re1}
\beta^{t_0}=-2t_0-\ap\sum_i u_i
\,\,,\quad
\beta^{w_i}=-2w_i
\,\,,\quad
\beta^{u_i}=-2u_i
\,\,,\quad
\beta^{\varphi_0}=\oh\lambda
\,\,,\quad
\beta^{v_i}=0
\,\,.
\end{equation}
Note that these expressions can be obtained by substituting the profiles \eqref{back}, with the bare 
couplings replaced by the renormalized ones, into the corresponding formulae \eqref{beta-1}. 

Finally, the renormalized partition function takes the form

\begin{equation}\label{2pf1}
Z_0=z_0\,\tilde V_d\,
\ep^{-2\varphi_0-t_0}
\prod_i \frac{1}{\sqrt{u_i}}\,
\ep^{\frac{s_i^2}{u_i}
+\oh\ap\bigl(1-\gamma\bigr)u_i+
\oh\sum_{m=2}^\infty \frac{(-1)^m}{m}\,\bigl(\ap u_i\bigr)^m
\sum^\infty_{n=1}\frac{2n+1}{(n^2+n)^m}}
\quad,
\end{equation}
with $s_i=w_i+v_i$.

In fact, the renormalization as it has been done in above is incomplete. The missing point is the renormalization 
conditions which fix the finite part of $Z_0$ removing ambiguities due to a particular renormalization scheme. 
So, a general form of $Z_0$ turns out to be  

\begin{equation}\label{2pf2}
Z_0=z_0\,\tilde V_d\,
\ep^{-2\varphi_0-t_0}
\prod_i \frac{1}{\sqrt{u_i}}\,
\ep^{\frac{s_i^2}{u_i}
+c_2\ap u_i+
\oh\sum_{m=2}^\infty \frac{(-1)^m}{m}\,\bigl(\ap u_i\bigr)^m
\sum^\infty_{n=1}\frac{2n+1}{(n^2+n)^m}}
\quad,
\end{equation}
where $c_2$ is an arbitrary coefficient. For the following, we set $c_2=0$ (see also \eqref{pf-ap1}).

To produce the Weyl anomaly coefficients, we substitute the profiles \eqref{back} into the corresponding 
expressions \eqref{Weyl}. As a result, 

\begin{equation}\label{weyl}
\bar\beta^{t_0}=\beta^{t_0}+2\ap\sum_iv_iw_i
\,\,,\quad
\bar\beta^{w_i}=\beta^{w_i}+\ap\,u_iv_i
\,\,,\quad
\bar\beta^{u_i}=\beta^{u_i}
\,\,,\quad
\bar\beta^{\varphi_0}=\beta^{\varphi_0}+\ap\sum_iv_i^2
\,\,,\quad
\bar\beta^{v_i}=\beta^{v_i}
\,\,.
\end{equation}
As we have noted in section 3.2, $\bar\beta^i\cdot\frac{\delta Z_0}{\delta g^i}=
\beta^i\cdot\frac{\delta Z_0}{\delta g^i}$. One can verify that this again holds. Plugging $Z_0$ and the 
$\beta$'s into \eqref{ac4} we then find

\begin{equation}\label{2pf3}
S_0=\Bigl(2+2t_0-\lambda+\ap\,\sum_i u_i-2\sum_i u_i\frac{\pd}{\pd u_i}-
4\sum_i\frac{1}{u_i}(w_i^2+v_iw_i)
\Bigr)Z_0+f_0
\quad.
\end{equation}

As a check, one can verify that modulo $f_0$ and its counterpart in \eqref{ap-a} the $\ap$-expansion as 
it follows from \eqref{2pf3} 

\begin{equation}\label{2pf4}
S_0=z_0\,\tilde V_d\,
\ep^{-2\varphi_0-t_0}
\Bigl(2+2t_0-\lambda+d+2\sum_i\frac{1}{u_i}(v_i^2-w_i^2)+\ap\sum_iu_i\Bigr)
\prod_i\frac{1}{\sqrt{u_i}}\,\ep^{\frac{s_i^2}{u_i}}\,\,
+O(\ap\,^2)
\end{equation}
coincides with the $\ap$-expansion obtained from Eq.\eqref{ap-a} by evaluating the action on 
the profiles \eqref{back}.\footnote{In doing so, it is advantageous to first rewrite the kinetic terms of the 
scalar fields in \eqref{ap-a} by integrating by parts 
$\int d^Dx\,\sqrt{G}\ep^{-2\varphi-t}\Bigl(4\pd\varphi\pd\varphi+4\pd\varphi\pd t+\pd t\pd t\Bigr)=
\int d^Dx\,\sqrt{G}\ep^{-2\varphi-t}\Bigl(2D^2\varphi+D^2t\Bigr)$.} It is straightforward to 
generalize the above analysis to include the effects of $f_0$ by modifying the dilaton profile 
to $\varphi(X)+\sum_i q_iX_i^2$. Note that in the case of $t=0$ the last term plays the role of a 
regulator for integrations over the $x$'s. 

It is worth noting that for $T=0$ the action derived from the p-adic string theory in section 2 has extrema 
which look like the quadratic profiles \eqref{back}. It was also one of the motivation for us to consider the free 
field background. As known, in the case of open bosonic string the quadratic profiles play a prominent role 
being identified with D-branes. Unfortunately, we don't have something similar to say on their role, if any, 
in the case of closed bosonic string. Note, however, that in the context of the p-adic string theory these 
solutions were interpreted as spacetimes of lower dimensionality related to noncritical p-adic 
strings \cite{ms}. Certainly, this issue deserves to be further clarified.

\subsection{Adding the open string tachyon}

It would be interesting to extend our analysis to other topologies. Some relevant methods to deal with 
the sigma model on a general Riemann surface were discussed in the literature (see \cite{ts2} for a review) but 
they are still insufficient. However, next topology to be considered - the disk - may be treated relatively easy. 
In the case of the simplest closed string background ($t=0,\,\lambda=0,\,G_{\mu\nu}=\eta_{\mu\nu}$) 
this has already been done in the framework of the background independent open string theory 
\cite{W,GS,KMM} and the sigma model approach \cite{ts1}. In this subsection our goal will be to extend this 
analysis to the case of constant closed string backgrounds.

Before continuing our discussion, we will make a short detour and recall some basic facts on the sigma model 
approach. First, the sigma model action has boundary terms

\begin{equation}\label{bound}
I_b=\frac{1}{2\pi}\oint dl\,\bigl[ \mathbf{T}(X)+K\boldsymbol{\varphi}(X)\bigr]
\quad.
\end{equation}
Second, the sigma model path integral is modified to include an integration over some additional 
parameters (moduli). Their role is to enhance $SL(2,R)$ symmetry to $SL(2,C)$ that allows one to treat 
both world-sheets on equal footing. One way to implement this is to consider the world-sheet as a sphere 
with a hole. Then a radius of the hole and its location play the role of moduli. Third, small hole divergences are 
canceled by a proper redefinition of the dilaton and graviton backgrounds. This leads to additional 
contributions to their $\beta$-functions \cite{fs}

\begin{equation}\label{fs}
\Delta\beta^\varphi=\qr N_D\Bigl(\oh D+1\Bigr)\ep^\varphi
\quad,\quad
\Delta\beta^G_{\mu\nu}=\oh N_D \,\ep^\varphi G_{\mu\nu}
\quad.
\end{equation}
Here $N_D$ is a numerical factor. We set $N_D=\kappa_0^2\ap T_{D-1}$, where $T_p$ is the 
tension of the $D_p$-brane. It is important for what follows that the tachyon $\beta$-function remains 
unmodified.

It is not hard to write down the renormalized partition function. We claim that it is 

\begin{equation}\label{dpf}
Z_\oh=T_{D-1}\int d^Dx\,\sqrt{G}\ep^{-\varphi-\oh t-T}
\quad.
\end{equation}

All features of \eqref{dpf} can be understood from general reasoning. Considering first just the open string 
tachyon background, the partition function reduces to that of the background independent open string theory. 
The dependence $\ep^{-\varphi-\oh t}$ arises because the effective coupling constant is given by
$g_{\text{eff}}=\ep^{\varphi+\oh t}$. $Z_0$ from the spherical topology is weighted by 
$g_{\text{eff}}^{-2}$, so $Z_\oh$ should be weighted by $g_{\text{eff}}^{-1}$. Finally, the volume factor 
$\int d^Dx\,\sqrt{G}$ follows from the integration over zero modes and general covariance.

The action can be determined as follows. We start by gathering the partition function $Z=Z_0+Z_\oh$. This 
results in 

\begin{alignat*}{2}
{\cal G}_{tt}&=\frac{1}{\kappa_0^2\ap} 
\Bigl(g_{\text{eff}}^{-2}+\qr N_D\,g_{\text{eff}}^{-1}\ep^{-T}\Bigr)V_D
\quad,&\qquad
{\cal G}_{t\varphi}&=2\,{\cal G}_{tt}
\quad,\\
{\cal G}_{tG}^{\mu\nu}&=-\frac{1}{2\kappa_0^2\ap}
\Bigl(g_{\text{eff}}^{-2}+\oh N_D \,g_{\text{eff}}^{-1}\ep^{-T}\Bigr)V_DG^{\mu\nu}
\quad,&\qquad
{\cal G}_{tT}&=\oh T_{D-1}V_D\,g_{\text{eff}}^{-1}\ep^{-T}
\quad.
\end{alignat*}
Plugging into Eq.\eqref{dif-c} with \footnote{In \eqref{fs} the field-dependent factor $\ep^\varphi$ arises from 
a ratio of the weights of the two topologies. In the presence of constant tachyon backgrounds this ratio picks up a 
factor $\ep^{\oh t-T}$, so the expressions \eqref{fs} get modified.}

\begin{equation}\label{bet}
\beta^t=-2t
\quad,\quad
\beta^\varphi=\oh\lambda+\qr N_D\Bigl(\oh D+1\Bigr)g_{\text{eff}}\,\ep^{-T}
\quad,\quad
\beta^G_{\mu\nu}=\oh N_D\,g_{\text{eff}}\,\ep^{-T}G_{\mu\nu}
\quad,\quad
\beta^T=-T
\end{equation}
we then find

\begin{equation}\label{acd}
S=S_0+T_{D-1}\int d^Dx\,\sqrt{G}\ep^{-\varphi}\biggl[\ep^{-\oh t-T}\bigl(1+t+T-\oh\lambda\bigr)
+f_\oh(\lambda,\varphi, T)\biggr]
\quad,
\end{equation}
where $S_0$ is given by \eqref{ac3}. $f_\oh$ is fixed by demanding that for $t=T=0$ the 
leading correction to $S_0$ is given by $T_{D-1}\int d^Dx\,\sqrt{G}(1+\oh\lambda)$. So, we 
find $f_\oh =\lambda$. As a check, one can verify that for 
$t=0,\,\lambda=0,\,G_{\mu\nu}=\eta_{\mu\nu}$ the second term reduces to the known result of 
\cite{W,GS,KMM}. Finally, the effective action is written as 

\begin{equation}\label{acd1}
\begin{split}
S=&\frac{2}{\kappa_0^2\ap}\int d^Dx\,\sqrt{G}\ep^{-2\varphi}
\biggl[\ep^{-t}\bigl(1+t-\oh\lambda\bigr)+\lambda-1\biggr]
\\
&+T_{D-1}\int d^Dx\,\sqrt{G}\ep^{-\varphi}\biggl[\ep^{-\oh t-T}\bigl(1+t+T-\oh\lambda\bigr)
+\lambda\biggr]
\quad.
\end{split}
\end{equation}

Now let us look a little more closely at this action.

(a) Shifting the tachyons $t\rightarrow t+\oh\lambda$, $T\rightarrow T-\oh t$ and setting 
$\varphi=\varphi_0,\,G_{\mu\nu}=\eta_{\mu\nu},\,g_c^2=\oh\kappa^2\ap,\,g_o^2=T_{D-1}^{-1}$, 
we find that modulo the last term in \eqref{T-ac} the two potentials coincide. This provides some 
evidence in favor of the universality of our results.

(b) The last term in \eqref{T-ac} is notable. It is responsible for the vanishing of the closed string tachyon 
tadpole near $t=0$. From this point of view the action \eqref{T-ac} as it follows from the p-adic string 
theory is not acceptable. So, we should take all conclusions drawn from its analysis with precaution. In 
contrast, the action \eqref{acd1} has the tadpole. One point concerning the tachyon tadpole worth of note: it 
seems that it is consistent with the dilaton and the graviton tadpoles. The story here is relatively simpler than 
in the famous computation of the $D$-brane tension \cite{Jb}. The relevant kinetic terms from \eqref{a26-3} 
or, even, \eqref{a26-2} are 

\begin{equation}\label{kin}
S=\frac{2}{\kappa_0^2\ap}\int d^{26}x\, \Bigl[\pd_\mu t\pd^\mu t+m^2\,t^2\Bigr]
\end{equation}
while the tadpole from \eqref{acd1} with $\lambda=0$ is 
\begin{equation}\label{tad}
S_{D-1}=\oh T_{D-1}\int d^Dx\,t
\quad.
\end{equation}
A simple field theory calculation of the tachyon exchange between two parallel D-branes separated by 
distance $R$ results in 

\begin{equation}\label{ex}
iV_D\frac{\pi}{2^{10}}\bigl(4\pi^2\ap\bigr)^{12-D}G_{26-D}(m^2,R)
\quad,
\end{equation}
where $G_p(m^2,R)$ means the propagator of a scalar particle of mass $m$ in $p$ dimensions. This is 
in agreement with the stringy calculation (see Eq.(8.7.17) in \cite{Jb}), where it is dominant in the closed 
string channel. 
 
(c) It is now high time to revise our analysis of section 2.2. According to \eqref{acd1}, the refined potential is 
\footnote{For convenience, we use the same coordinates $(\phi,\Phi)$.}

\begin{equation}\label{T-p1}
V=-\frac{1}{4g^2_c}\,\phi^2\ln\frac{\phi^2}{\ep}-
\frac{1}{4g^2_o}\,\Phi^2
\biggl(\ln\frac{\Phi^2}{\ep}+\oh\ln\phi^2\biggr)
\quad.
\end{equation}
A simple calculation shows that the potential is stationary at three points of the 
$\phi\text{\small -}\Phi$ plane: 

(i) $\phi=\phi_0,\,\Phi=1/\sqrt{\phi_0}$, where $\phi_0$ is a solution to $2\phi^3\ln\phi^2=-g$. It is 
a local maximum which slightly gets shifted from that of section 2. Indeed, we have 
$\phi_0=\sum c_ng^n=1-\qr g+O(g^2)$. 

(ii) $\phi=1,\,\Phi=0$. This is a saddle point. Excitations of $\Phi$ are completely suppressed while 
excitations of $\phi$ develop instability. As in section 2, we interpret this point as a closed string background 
(unstable) without D-branes. We may think of the rolling from the local maximum to this point as the 
process of open string tachyon condensation. The closed string tachyon again suffers from the process 
relatively slightly, e.g., its vacuum expectation value gets shifted: $\langle\phi\rangle =1-\qr g+O(g^2)
\rightarrow \langle\phi\rangle =1.$

(iii) $\phi=0,\,\Phi=0$.\footnote{Note that the potential is ambiguous as $\phi\rightarrow 0,\,
\Phi\rightarrow 0$. We use the following prescription to fix it. We introduce the polar coordinates 
$(r,\theta)$ and then set $r^2\ln\theta^2=0$ as $r\rightarrow 0,\,\theta\rightarrow 0$.} This point is 
notable. It is a local minimum. Both kinds of excitations are suppressed. It is unclear whether this point may be 
interpreted as a true vacuum. The problem is again that the potential is not bounded from below. 

A final interesting remark: it follows from our analysis that the closed string tachyon tadpoles which are
of course due to D-branes might be crucial for the theory to have a local vacuum. 

\section{Concluding comments} 
\renewcommand{\theequation}{4.\arabic{equation}}
\setcounter{equation}{0}

\subsection{Type 0 theories}
Having discussed the closed and open-closed bosonic string theories, we will try to apply this technique to 
type 0 theories. In this paper our goal is to calculate the potential of the tachyon from the 
$(NS-,NS-)$ sector. So, we do not distinguish type 0A from type 0B or vice versa.

Type 0 strings are described by $N=1$ supersymmetric world-sheet theories. Thus, what we need is to 
supersymmetrize the sigma model action \eqref{sigma}. For the sake of simplicity, we set $G_{\mu\nu}=
\eta_{\mu\nu}$ for a moment. A naive attempt to use 
$\int d^2zd^2\theta \, t(Y)$, with $Y^\mu=X^\mu+\theta\psi^\mu+\bar\theta\bar\psi^\mu$, 
fails. Indeed, it leads to the world-sheet interaction $\int d^2z \,\psi^\mu\bar\psi^\nu\pd_{\mu\nu} t$, so 
the partition function depends on $t$ only via its derivatives. As a result, it seems that there is no tachyon 
potential in the effective action. This is wrong because at least near the standard perturbative vacuum it 
must be $V\sim t^2$ to reproduce the tachyon mass. In the case of open superstring there was a similar 
problem because a naive use of $\oint dld\theta\,T(Y)$ also failed to reproduce non-vanishing tachyon 
potential. There are two possibilities to overcome it \cite{sW,a4}. We will proceed in a similar way.

One possibility to do so is the following \cite{a4}. Taking the simplest quadratic profile 
$t(X)=uX^2$ of subsection 3.4 which is a solution of the effective action in the 
bosonic case and supersymmetrizing it \footnote{We set $t_0=0$ because it is irrelevant for SUSY.} 

\begin{equation}\label{s-t}
\int d^2z \,u\Bigl(X^2-\ap\psi\pd^{-1}\psi-\ap\bar\psi\bar\pd^{-1}\bar\psi\Bigr)
\end{equation}
we immediately get the almost desired result if we assume that the corresponding ``supersymmetric'' 
tachyon profile is linear \footnote{We thank D. Sorokin for a discussion of this issue.}

\begin{equation}\label{s-t1}
\int d^2z \Bigl[t^2-\ap\pd_xt\psi\pd^{-1}\pd_xt\psi-\ap\pd_xt\bar\psi\bar\pd^{-1}\pd_xt\bar\psi+
O(\pd^2_xt)\Bigr]
\quad.
\end{equation}
Here $\pd=\pd/\pd z,\,\bar\pd=\pd/\pd\bar z$, and $\pd_x=\pd/\pd X$. The linear profile is given by 
$t(X)=aX$, with $u=a^2$.

Another possibility is to generalize the proposal of \cite{sW} to the problem of interest. At this point let us remind 
that in the case of open superstring the sigma model action is modified by introducing an auxiliary spinor 
superfield $\Gamma$. The relevant terms are 

\begin{equation}\label{s-t2}
\oint dld\theta\,\Bigl[\Gamma D\Gamma+T(Y)\Gamma\Bigr]
\quad.
\end{equation}
The desired result is obtained by integrating $\Gamma$ out. 

So, we introduce an auxiliary scalar superfield $Q$ whose expansion is $Q=Z+\theta\xi+
\bar\theta\bar\xi+\bar\theta\theta F$. We then modify the world-sheet action by 

\begin{equation}\label{s-t4}
\int d^2z d^2\theta\,\Bigl[ DQ\bar D Q+t(Y)Q\Bigr]
\quad.
\end{equation}
As before, the auxiliary field is easily integrated out. The result is 

\begin{equation}\label{s-t5}
\qr\int d^2z \,\Bigl[t^2-\ap\pd_\mu t\psi^\mu\pd^{-1}\pd_\nu t\psi^\nu-
\ap\pd_\mu t\bar\psi^\mu\bar\pd^{-1}\pd_\nu t\bar\psi^\nu
 +\ap{}^2\pd_{\mu\nu} t\psi^\mu\bar\psi^\nu(\pd\bar\pd)^{-1}
\pd_{\sigma\rho} t\psi^\sigma\bar\psi^\rho\Bigr]
\end{equation}
that coincides with \eqref{s-t1} modulo the $\pd^2t$-term which is irrelevant in our approximation.

Now, let us analyze the partition function. In the case of constant backgrounds, the analysis is simple. 
Since we are interested in the $(NS-,NS-)$ sector, it goes along the lines of subsection 3.1. For $D=10$ 
the renormalized partition function is 

\begin{equation}\label{s-t3} 
Z_0=z_0\int d^{10}x\,\sqrt{G}\,\ep^{-2\varphi-\qr t^2}
\quad.
\end{equation}

As long as $S_\oh=Z_\oh$ modulo terms which are independent of the tachyons, it seems natural 
to suggest that in our approximation (constant backgrounds) 
$S_0=Z_0+f_0$, where $f_0$ is independent of $t$. As in section 3, $f_0$ is fixed by demanding that for $t=0$ 
the action reduces to the known result of \cite{Jb2}. We have

\begin{equation}\label{s-t6} 
S_0=\frac{2}{\kappa_0^2\ap}\int d^{10}x\,\sqrt{G}\,\ep^{-2\varphi}\bigl[\ep^{-\qr t^2}-1\bigr]
\quad.
\end{equation}
Here we set $z_0=2/\kappa_0^2\ap$. Alternatively, the action can be derived by solving Eq.\eqref{dif-c} for 
$\beta^t=-t$ and ${\cal G}_{tt}=\oh Z_0$. The form of ${\cal G}_{tt}$ follows from the discussion of open 
superstring, where the tachyon potential has the same form $V=\ep^{-\qr T^2}$. Note that the potential is 
an even function of $t$. This is consistent with the fact that the tachyon amplitudes with odd numbers of 
external legs vanish. 

The potential following from \eqref{s-t6} is stationary at: (i) $t=0$. This is a local maximum which is 
the standard perturbative vacuum of the theory. (ii) $t=\infty$. This is a vacuum to which the tachyon 
condenses. Unlike the bosonic case the potential is bounded from below that makes the process of condensation 
more peaceful as expected in \cite{am}.

\subsection{Many open problems}

There is a large number of open problems associated with the circle of ideas explored in the paper. In this 
subsection we list a few problems, perhaps, the most interesting ones.

In the bosonic case our analysis of subsection 3.5 revealed the three stationary points of the potential. The 
rolling from (i) to (ii) has attracted much attention as the process of open string tachyon condensation. 
Usually, the point (ii) is interpreted as a closed string vacuum. It follows from our analysis that this is a 
saddle point rather than the endpoint of the process. The closed string tachyon is responsible for developing a 
perturbative instability (one-loop effect) which will drive the theory to (iii) \footnote{In this paper we don't 
consider the possibility for the closed tachyon instead to roll down to $\Phi=\infty$.}. Note that there is still 
some possibility for tunneling through the potential barrier but its decay rate per unit volume is much 
smaller in the weak coupling regime.  The novelty is that the system can roll down directly from (i) to 
(iii) which is a local minimum. It seems natural to interpret this as open-closed tachyon condensation. If so, 
then it would be very interesting to understand the endpoint of this process, i.e., the point (iii). Some conclusions 
we made in subsection 3.3. might be useful in doing so. In particular, some of the closed string modes survive 
and the dilaton develops a linear profile. It is not clear however whether the latter is a hint on non-critical 
string theory at this vacuum.

One of our observation is that the closed string tachyon tadpole or, equivalently, unstable D-brane turned out 
to be crucial for the system to have a local minimum. The situation is more complicated in type 0 theories, 
where in addition to unstable D-branes there are unstable ${\text D}\bar{\text D}$ systems. Although some 
points of our analysis can be easily generalized to type 0 theories, there is still a lot of work ahead. It would be 
nice to make further progress along the lines of \cite{am,kts}.

Another related issue is the terms associated with the tachyons in the NSR formulation of the 
string sigma model. The appearance of the auxiliary fields do not seem quite natural from the viewpoint of the 
string path integral. This is perhaps a cost for the use of the ghost number zero picture. If so, then the 
experience with the tachyon backgrounds might be useful to understand the long standing problem of the 
NSR formulation: the RR backgrounds. The issue is worth being further clarified.

Finally, as we have emphasized at various points, the closed and open tachyon potentials have the same form. 
It world be very interesting to understand whether this observation is universal or model-dependent.

\vspace{.25cm} {\bf Acknowledgments}

\vspace{.25cm} 
We wish to thank I. Bars for encouragement, and H. Dorn and A.A. Tseytlin for comments on the 
manuscript. We gratefully acknowledge the hospitality of the Institute for Pure and 
Applied Mathematics at UCLA, where some of this work was carried out. The work 
is supported in part by DFG under Grant No. DO 447/3-1 and the European Commission 
RTN Programme HPRN-CT-2000-00131.


\end{document}